\documentclass{elsart}

\usepackage{graphicx}

\newcommand{\ACP}{A_{CP}}
\newcommand{\BB}{B\overline{B}}
\newcommand{\BF}{\mathcal{B}}
\newcommand{\bsgamma}{b \to s\gamma}
\newcommand{\CP}{\mathit{CP}}

\newcommand{\coshel}{\cos\theta_\mathrm{hel}}
\newcommand{\DE}{\Delta E}
\newcommand{\Ebeam}{E_{\mathrm{beam}}^{*}}
\newcommand{\Egamma}{E_\gamma^*}
\newcommand{\fbi}{\mathrm{fb}^{-1}}
\newcommand{\GeV}{\mathrm{GeV}}
\newcommand{\Ktwost}{K_2^*(1430)}
\newcommand{\Kthreest}{K_3^*(1780)}

\newcommand{\Mbc}{M_{\mathrm{bc}}}
\newcommand{\MeV}{\mathrm{MeV}}
\newcommand{\MKeta}{M_{K\eta}}
\newcommand{\KS}{K^0_S}
\newcommand{\qq}{q\overline{q}}
\newcommand{\thetahel}{\theta_\mathrm{hel}}

\newcommand{\PM}[2]{{\,}^{+#1}_{-#2}}

\newcommand{\LumONRES}{253}
\newcommand{\LumOFFRES}{28}
\newcommand{\NBBmillunit}{275}
\newcommand{\SYSTNBBstr}{1.1\%}

\newcommand{\BFbsgammaPDG}{(3.3 \pm 0.4) \times 10^{-4}}

\newcommand{\effMxsforbsgamma}{84\%}
\newcommand{\effKID}{90\%}
\newcommand{\fakeKID}{10\%}
\newcommand{\effPIID}{98\%}

\newcommand{\MCeffLRsig}{44\%}
\newcommand{\MCrejLRqq}{98\%}

\newcommand{\YieldCharged}{81 \pm 14 \PM{10}{6}}
\newcommand{\YieldNeutral}{20.9 \PM{7.3}{6.5} \PM{4.2}{3.2}}
\newcommand{\YieldTotal}{102 \pm 16 \PM{13}{8}}
\newcommand{\YieldTc}{4.4 \PM{5.2}{4.5} \PM{2.6}{2.4}}
\newcommand{\YieldTn}{0.2 \PM{3.1}{2.4} \PM{1.3}{1.4}}
\newcommand{\YieldTt}{5.2 \PM{5.9}{5.2} \PM{3.5}{3.2}}
\newcommand{\YieldULTc}{15.0}
\newcommand{\YieldULTn}{7.5}
\newcommand{\YieldULTt}{17.7}

\newcommand{\YieldChargedNosyst}{81 \pm 14}
\newcommand{\YieldNeutralNosyst}{20.9 \PM{7.3}{6.5}}
\newcommand{\YieldTotalNosyst}{102 \pm 16}

\newcommand{\SNFstatCharged}{7.1}
\newcommand{\SNFstatNeutral}{3.7}
\newcommand{\SNFstatTotal}{8.1}
\newcommand{\SNFsystCharged}{6.8}
\newcommand{\SNFsystNeutral}{3.4}
\newcommand{\SNFsystTotal}{7.7}

\newcommand{\EFFc}{3.50 \pm 0.27}
\newcommand{\EFFn}{0.87 \pm 0.08}
\newcommand{\EFFt}{4.37 \pm 0.31}
\newcommand{\EFFTc}{2.03 \pm 0.16}
\newcommand{\EFFTn}{0.48 \pm 0.05}
\newcommand{\EFFTt}{2.51 \pm 0.18}

\newcommand{\SYSTphotonSTR}{2.8\%}
\newcommand{\SYSTkaonidSTR}{0.8\%}
\newcommand{\SYSTpionidSTR}{0.5\%}
\newcommand{\SYSTksSTR}{4.5\%}
\newcommand{\SYSTpizeroSTR}{1.5\%}
\newcommand{\SYSTetaSTR}{2.0\%}
\newcommand{\SYSTlrsystCstr}{5.9\%}
\newcommand{\SYSTlrsystNstr}{4.4\%}
\newcommand{\SYSTketamassCstr}{2.1\%}
\newcommand{\SYSTketamassNstr}{4.4\%}
\newcommand{\SYSTcoshelCstr}{2.5\%}
\newcommand{\SYSTcoshelNstr}{3.4\%}
\newcommand{\SYSTsubbfgSTR}{0.7\%}
\newcommand{\SYSTsubbfpSTR}{1.8\%}

\newcommand{\BFketagammaZc}{8.4 \pm 1.5 \PM{1.2}{0.9}}
\newcommand{\BFketagammaZn}{8.7 \PM{3.1}{2.7} \PM{1.9}{1.6}}
\newcommand{\BFketagammaZt}{8.5 \pm 1.3 \PM{1.2}{0.9}}
\newcommand{\BFkthreestgammaULc}{2.9}
\newcommand{\BFkthreestgammaULn}{6.4}
\newcommand{\BFkthreestgammaULt}{2.8}
\newcommand{\BFkthreestgammaULcSBstr}{3.9 \times 10^{-5}}
\newcommand{\BFkthreestgammaULnSBstr}{8.3 \times 10^{-5}}
\newcommand{\BFkthreestgammaULtSBstr}{3.7 \times 10^{-5}}

\newcommand{\BFketagammaZcSTR}{( 8.4 \pm 1.5 \mbox{(stat)}%
 \PM{1.2}{0.9} \mbox{(syst)} ) \times 10^{-6}}

\newcommand{\AcpSystFit}{0.045}
\newcommand{\AcpSystDet}{0.035}
\newcommand{\AcpSystPID}{0.014}

\newcommand{\YieldZcMinus}{34.0 \PM{9.8}{9.0}}
\newcommand{\YieldZcPlus}{46.7 \PM{10.5}{9.8}}

\newcommand{\AcpZc}{-0.16 \pm 0.09 \pm 0.06}

\begin{document}

\begin{frontmatter}



\vspace*{-3\baselineskip}
\begin{flushleft}
 \resizebox{!}{3cm}{\includegraphics{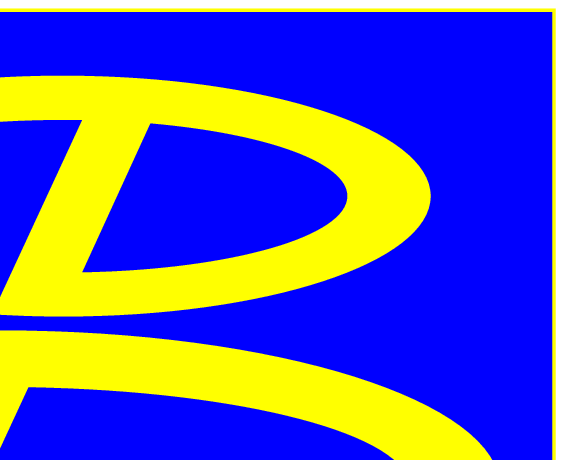}}
\end{flushleft}
\vspace*{-3cm}
\begin{flushright}
 Belle Preprint 2004-34 \\
 KEK Preprint 2004-69
\end{flushright}
\vspace*{2cm}

\title{Observation of $B^+ \to K^+\eta\gamma$}


\collab{Belle Collaboration}
  \author[KEK]{S.~Nishida}, 
  \author[KEK]{K.~Abe}, 
  \author[Tokyo]{H.~Aihara}, 
  \author[Nagoya]{M.~Akatsu}, 
  \author[Tsukuba]{Y.~Asano}, 
  \author[BINP]{V.~Aulchenko}, 
  \author[ITEP]{T.~Aushev}, 
  \author[Cincinnati]{S.~Bahinipati}, 
  \author[Sydney]{A.~M.~Bakich}, 
  \author[Peking]{Y.~Ban}, 
% ...

\address[BINP]{Budker Institute of Nuclear Physics, Novosibirsk, Russia}
% ...
\address[Cincinnati]{University of Cincinnati, Cincinnati, OH, USA}
% ...
\address[KEK]{High Energy Accelerator Research Organization (KEK), Tsukuba, Japan}
% ...
\address[ITEP]{Institute for Theoretical and Experimental Physics, Moscow, Russia}
% ...
\address[Nagoya]{Nagoya University, Nagoya, Japan}
% ...
\address[Peking]{Peking University, Beijing, PR China}
% ...
\address[Sydney]{University of Sydney, Sydney, NSW, Australia}
% ...
\address[Tokyo]{Department of Physics, University of Tokyo, Tokyo, Japan}
% ...
\address[Tsukuba]{University of Tsukuba, Tsukuba, Japan}
% ...
\thanks[NovaGorica]{on leave from Nova Gorica Polytechnic, Nova Gorica, Slovenia}


\begin{abstract}
 We report measurements of radiative $B$ decays with $K\eta\gamma$
 final states, using a data sample of $\LumONRES~\fbi$ recorded at the
 $\Upsilon(4S)$ resonance with the Belle detector at the KEKB $e^+e^-$
 storage ring.
 We observe $B^+ \to K^+\eta\gamma$ for the first time
 with a branching fraction of $\BFketagammaZcSTR$ for
 $\MKeta < 2.4~\GeV/c^2$,
 and find evidence of $B^0 \to K^0\eta\gamma$.
 We also search for $B \to \Kthreest\gamma$.
\end{abstract}

\end{frontmatter}


Radiative $B$ decays,
which proceed mainly through the $\bsgamma$ process\footnote{%
Throughout this paper, the inclusion of the charge conjugate mode
is implied unless otherwise stated.},
have played an important role in a search for physics beyond
the Standard Model (SM).
Although the inclusive branching fraction has been measured
to be $\BFbsgammaPDG$~\cite{Eidelman:2004wy},
we know little about its constituents.
So far, measured exclusive final states such as
$K^*(892)\gamma$~\cite{Coan:1999kh,Aubert:2001me-Nakao:2004th},
$\Ktwost\gamma$~\cite{Coan:1999kh,Nishida:2002me},
$K\pi\pi\gamma$~\cite{Nishida:2002me} and
$K\phi\gamma$~\cite{Drutskoy:2003xh}
only explain one third of the inclusive rate.
Detailed knowledge of exclusive final states
reduces the theoretical uncertainty in the measurement
of the inclusive $B \to X_s\gamma$ branching fraction
using the pseudo-reconstruction technique,
as well as in the measurement of $B \to X_s\ell^+\ell^-$~\cite{Kaneko:2002mr}.
In this analysis, the decay mode $B \to K\eta\gamma$
is studied for the first time.
In addition to improving the understanding of $\bsgamma$ final states,
$B^0 \to \KS\eta\gamma$ can be used
to study time-dependent $\CP$ asymmetry~\cite{TCPV-bsgamma},
which is sensitive to physics beyond the SM.
The mode $B \to K\eta\gamma$ can also be used to search for
radiative $B$ decays through possible $K\eta$ resonances,
e.g., $\Kthreest$ observed by the LASS experiment~\cite{Aston:1987ey}.

The analysis is based on
$\LumONRES~\fbi$ of data taken at the $\Upsilon(4S)$ resonance
(on-resonance) and
$\LumOFFRES~\fbi$ at an energy $60~\MeV$ below the resonance
(off-resonance),
which were recorded by the Belle detector~\cite{Mori:2000cg}
at the KEKB asymmetric $e^+e^-$ collider
($3.5~\GeV$ on $8~\GeV$)~\cite{KEKB:NIM}.
The on-resonance data corresponds to $\NBBmillunit$ million $\BB$ events.
The Belle detector
is comprised of a silicon vertex detector,
a 50-layer central drift chamber (CDC), an array of aerogel
Cherenkov counters (ACC), time-of-flight scintillation counters (TOF)
and an electromagnetic calorimeter of CsI(Tl) crystals (ECL) located inside
a superconducting solenoid coil that provides a 1.5 T magnetic field.
An instrumented iron flux-return for $K_L^0$/$\mu$ detection
is located outside the coil.
Two different inner detector configurations were used.
For the first sample of 152 million $B\overline{B}$ pairs,
a 2.0 cm radius beampipe
and a 3-layer silicon vertex detector were used;
for the latter 123 million $B\overline{B}$ pairs,
a 1.5 cm radius beampipe, a 4-layer silicon detector
and a small-cell inner drift chamber were used~\cite{Ushiroda}.

We reconstruct $B^+ \to K^+\eta\gamma$ and $B^0 \to \KS\eta\gamma$
via $\eta \to \gamma\gamma$ and $\eta \to \pi^+\pi^-\pi^0$.
All charged tracks used in the reconstruction
(except charged pions from $\KS$)
are required to have a center-of-mass (CM)
momentum greater than $100~\MeV/c$
and to have an impact parameter
within $\pm 5 \mathrm{~cm}$
of the interaction point along the positron beam axis
and within $0.5 \mathrm{~cm}$ in the transverse plane.
In order to identify kaon and pion candidates,
we use a likelihood ratio based on the light yield in the ACC,
TOF information and specific ionization measurements in the CDC.
For the requirement applied on the likelihood ratio,
we obtain an efficiency
(pion misidentification probability) of $\effKID$ ($\fakeKID$)
for charged kaon candidates,
and an efficiency
(kaon misidentification probability) of $\effPIID$ ($\fakeKID$)
for charged pion candidates.
Tracks identified as electrons are excluded.

$\KS$ candidates are formed from $\pi^+\pi^-$ combinations
with invariant mass within $8~\MeV/c^2$ ($\sim 2\sigma$)
of the nominal $\KS$ mass.
The two pions are required to have a common vertex
displaced from the interaction point.
The $\KS$ momentum direction is required to be
consistent with the $\KS$ flight direction.
Neutral pion candidates are formed from pairs of photons
that have an invariant mass within $16~\MeV/c^2$ ($\sim 3\sigma$) of
the nominal $\pi^0$ mass and a momentum greater than $100~\MeV/c$
in the CM frame.
Each photon is required to have an energy greater than $50~\MeV$
in the laboratory frame.
A mass-constrained fit is then performed to obtain the $\pi^0$ momentum.

For $\eta \to \gamma\gamma$ reconstruction, we require that
the invariant mass of the two photons satisfy
$0.515~\GeV/c^2 < M_{\gamma\gamma} < 0.570~\GeV/c^2$
and that each photon have an energy greater than $50~\MeV$
in the laboratory frame.
We also require $|\coshel^{\eta}| < 0.9$,
where $\thetahel^\eta$ is the 
angle between the photon momentum and $\eta$ boost direction from 
the laboratory frame in the $\eta$ rest frame.
A mass-constrained fit is then performed to obtain the $\eta$ momentum.
For $\eta \to \pi^+\pi^-\pi^0$,
we apply a requirement
on the three-pion invariant mass,
$0.532~\GeV/c^2 < M_{\pi^+\pi^-\pi^0} < 0.562~\GeV/c^2$.

We reconstruct $B$ meson candidates from an $\eta$,
a charged or neutral kaon and the highest energy photon
within the acceptance of the barrel ECL
($33^\circ<\theta_\gamma<128^\circ$, where $\theta_\gamma$
is the polar angle of the photon 
with respect to the electron beam
in the laboratory frame).
Here, the invariant mass of the $K\eta$ system is required to be less
than $2.4~\GeV/c^2$.
This selection corresponds to $E_\gamma^{(B)} > 2.1~\GeV$,
where $E_\gamma^{(B)}$ is the photon energy in the $B$ rest frame,
and includes $\effMxsforbsgamma$ of events from the $\bsgamma$ process.
The highest energy photon candidate is required
to be consistent with an isolated electromagnetic shower,
i.e., 95\% of the energy in an array of $5 \times 5$ crystals
should be concentrated in
an array of $3 \times 3$ crystals
and no charged tracks should be associated with it.
In order to reduce the background from
decays of $\pi^0$ and $\eta$ mesons,
we combine the photon candidate
with each of the other photons
that have CM energy greater than $30~\MeV$ ($200~\MeV$)
in the event
and reject the event if the invariant mass of any pair is
within $18~\MeV/c^2$ ($32~\MeV/c^2$)
of the nominal $\pi^0$ ($\eta$) mass.
This condition is referred to as the $\pi^0/\eta$ veto.

We use two independent kinematic variables for the $B$ reconstruction:
the beam-energy constrained mass
$\Mbc \equiv \sqrt{\left(\Ebeam/c^2\right)^2
  - (|\vec{p}_{K\eta}^{\,*}+\vec{p}_\gamma^{\,*}|/c)^2}$
and
$\DE \equiv E_{K\eta}^* + \Egamma - \Ebeam$,
where $\Ebeam$ is the beam energy,
and $\vec{p}_\gamma^{\,*}$, $\Egamma$,
$\vec{p}_{K\eta}^{\,*}$, $E_{K\eta}^*$ are
the momenta and energies of the photon
and the $K\eta$ system, respectively, calculated in the CM frame.
In the $\Mbc$ calculation, the photon momentum is rescaled
so that $|\vec{p}_\gamma^{\,*}|=(\Ebeam-E_{K\eta}^*)/c$
is satisfied.
We require $\Mbc > 5.2~\GeV/c^2$ and $-150~\MeV < \DE < 80~\MeV$.
We define the $B$ signal region to be $\Mbc > 5.27~\GeV/c^2$.
In the case that multiple candidates are found in the same event,
we take the candidate that has the $\eta$ mass closest to the nominal
mass\footnote{In case multiple candidates share such an $\eta$ candidate,
the candidate with the smallest $|\DE|$ is chosen.}
after applying the background suppression described later.

The largest source of background originates from
continuum $e^+e^- \to \qq$ ($q = u,d,s,c$) production
including contributions from
initial state radiation ($e^+e^-\to \qq\gamma$).
In order to suppress this background,
we use the likelihood ratio (LR) described in Ref.~\cite{Nishida:2002me},
which utilizes the information from
a Fisher discriminant~\cite{Fisher:1936et}
formed from six modified Fox-Wolfram moments~\cite{Fox:1978vu}
and the cosine of the angle between
the $B$ meson flight direction and the beam axis.
The LR requirement
retains $\MCeffLRsig$ of the signal,
while rejecting $\MCrejLRqq$ of the continuum background.

In order to extract the signal yield, we perform
a binned likelihood fit to the $\Mbc$ distribution.
The $\Mbc$ distribution of the signal component is modeled
by a Crystal Ball line shape~\cite{Crystal-Ball},
with the parameters determined from the signal Monte Carlo (MC)
and calibrated using
control samples of $B^+ \to \overline{D}{}^0(\to K^+\pi^-\pi^0)\pi^+$ and
$B^0 \to D^-(\to \KS\pi^-\pi^0)\pi^+$ decays.
The $\Mbc$ distribution of the continuum background is
modeled by an ARGUS function~\cite{Albrecht:1990am}
whose shape is determined from the off-resonance data.
Here, the LR requirement is not applied to the off-resonance data
in order to compensate for the limited amount of data in that sample.
The possible bias due to this is taken as systematic error on the fitted
yield.
Background from hadronic $B$ decays is divided into two components, which
we refer to as generic $\BB$ background
and rare $B$ background in this paper.
The former comprises $B$ decays through $b \to c$ transitions including
color-suppressed $B$ decays such as $B^0 \to \overline{D}{}^0\pi^0$,
and the latter covers charmless $B$ decays.
Each of them is modeled by another ARGUS function.
The shape of these distributions is determined using
corresponding MC samples.
In order to study the contamination from other $\bsgamma$ decays,
we examine a $B \to K^*(892)\gamma$ MC sample
and an inclusive $\bsgamma$ MC sample that is
modeled as an equal mixture of $s\overline{d}$ and $s\overline{u}$
quark pairs and is hadronized using JETSET~\cite{Sjostrand:1994yb},
where the $X_s$ mass spectrum is fitted to the model of Kagan and
Neubert~\cite{Kagan:1998ym}.
We find that feed-down from other $\bsgamma$ decays is small,
but not negligible,
and model its $\Mbc$ distribution with an ARGUS function.

\begin{figure}
 \begin{center}
  \includegraphics[scale=0.62]{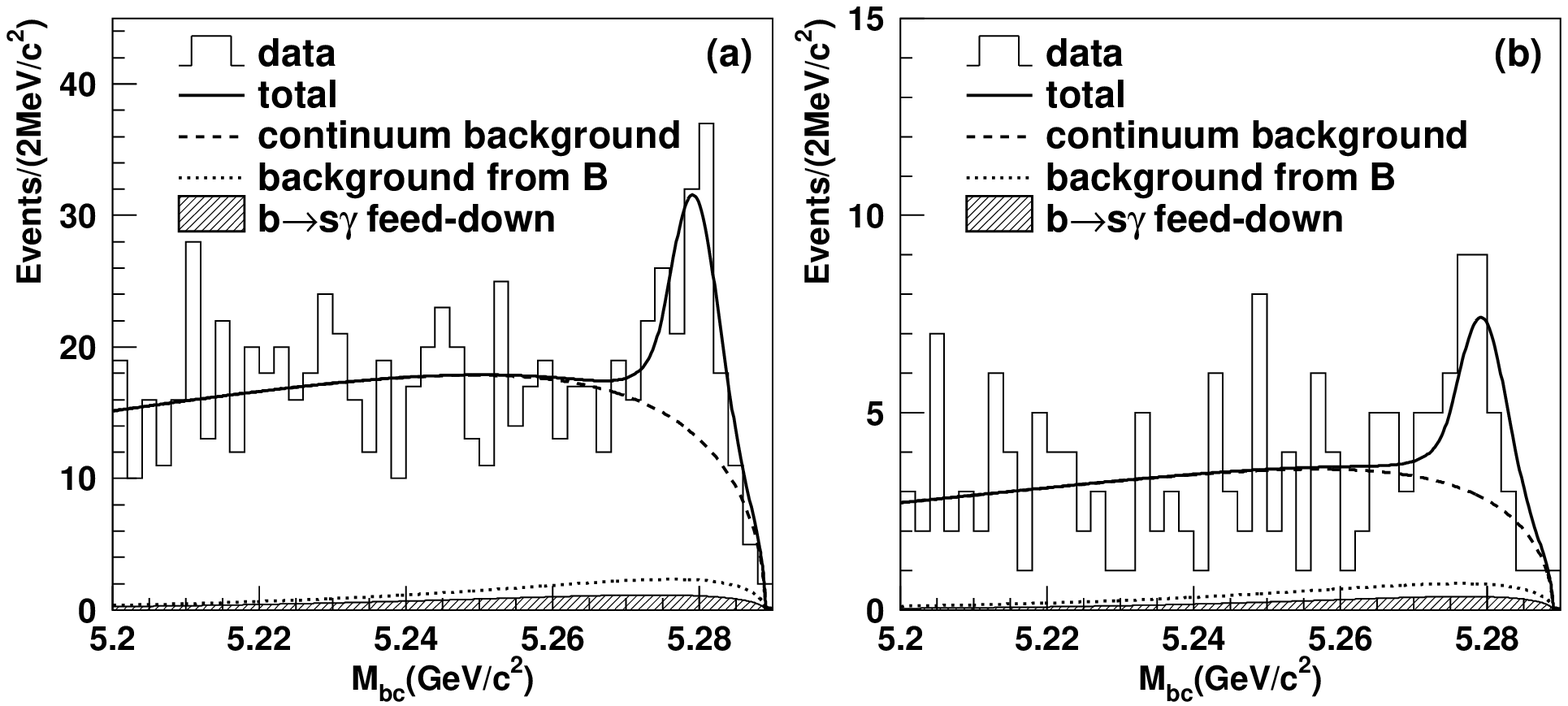}
  \caption{\label{ketagamma_fig1}%
  $\Mbc$ distributions for (a) $B^+ \to K^+ \eta\gamma$,
  (b) $B^0 \to \KS \eta\gamma$.
  Fit results are overlaid.}
 \end{center}
\end{figure}

Figure~\ref{ketagamma_fig1} shows the $\Mbc$ distributions
for $B^+ \to K^+ \eta\gamma$ and $B^0 \to \KS \eta\gamma$, respectively.
These distributions, as well as the distribution for
the combined mode, are fitted to the sum of signal, continuum,
generic $\BB$, rare $B$ background and $\bsgamma$ feed-down components.
In the fit, the normalization of generic $\BB$, rare $B$ and $\bsgamma$
are fixed according to the luminosity and $\bsgamma$ branching fraction,
while the normalization of the continuum component is allowed to float.
We find signal yields of $\YieldChargedNosyst$,
$\YieldNeutralNosyst$ and $\YieldTotalNosyst$ events with
statistical significances of
$\SNFstatCharged\sigma$, $\SNFstatNeutral\sigma$ and $\SNFstatTotal\sigma$,
for the charged, neutral and combined modes, respectively.
Here, the significance is defined as
$\sqrt{- 2 \ln ( \mathcal{L}(0) / \mathcal{L}_{\mathrm{max}} )}$, where
$\mathcal{L}_{\mathrm{max}}$ and $\mathcal{L}(0)$
are the maximum values of the likelihood
when the signal yield is left free or fixed to zero, respectively.

\begin{figure}
 \begin{center}
  \includegraphics[scale=0.62]{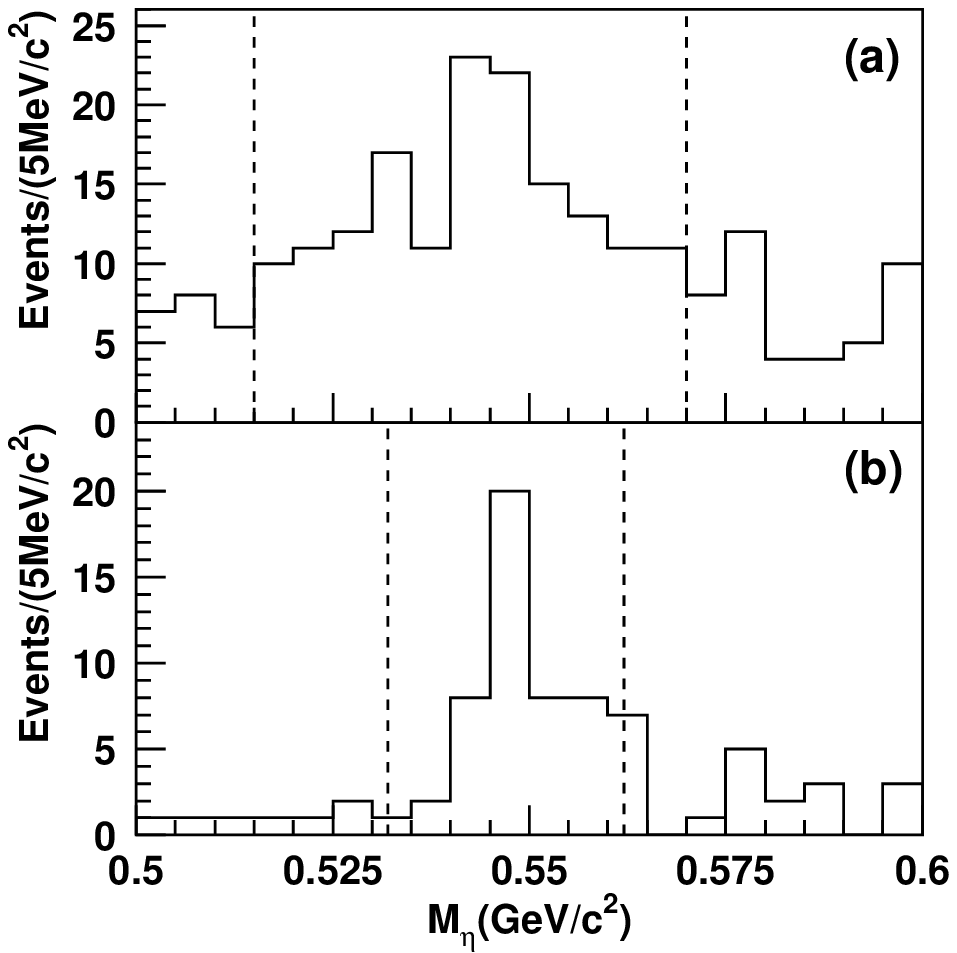}
  \caption{\label{ketagamma_fig2}%
  $\eta$ invariant mass distributions for (a) $\eta \to \gamma\gamma$
  and (b) $\eta \to \pi^+\pi^-\pi^0$   
  inside the $B$ signal region for combined $B \to K\eta\gamma$.
  Dashed lines show the selection applied in the analysis.}
 \end{center}
\end{figure}

\begin{figure}
 \begin{center}
  \includegraphics[scale=0.62]{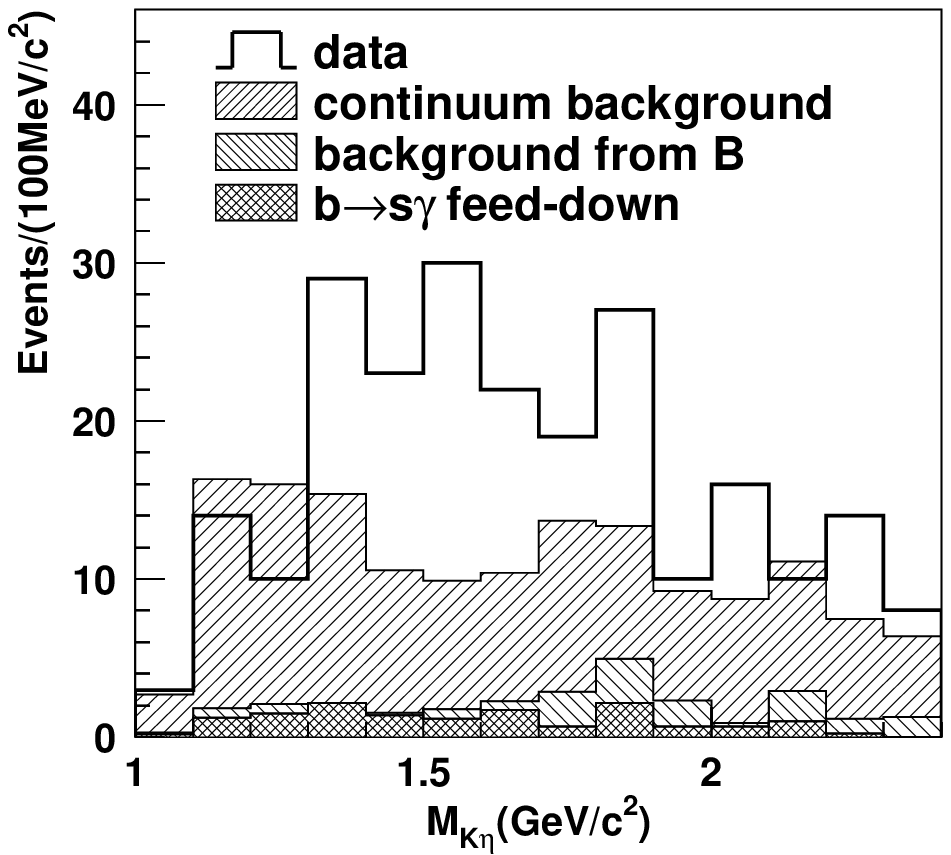}
  \caption{\label{ketagamma_fig3}%
  $K\eta$ invariant mass distribution for events
  inside the $B$ signal region for combined $B \to K\eta\gamma$.}
 \end{center}
\end{figure}

Figure~\ref{ketagamma_fig2} shows
the $\gamma\gamma$ and $\pi^+\pi^-\pi^0$ invariant
mass distributions for events inside the $B$ signal region.
Here, we do not apply the best candidate selection.
We observe clear peaks at the nominal $\eta$ mass.
The $K\eta$ invariant mass distribution for events
inside the $B$ signal region is shown in Fig.~\ref{ketagamma_fig3}.
Here, the background distributions are obtained from
the off-resonance data without the LR requirement
or from the corresponding MC samples,
and are normalized using the fit result.
We find that the signal
is concentrated between $1.3~\GeV/c^2$ and $1.9~\GeV/c^2$
and is falling above $1.9~\GeV/c^2$.
Therefore, our requirement $\MKeta < 2.4~\GeV/c^2$
is expected to include most of the $B \to K\eta\gamma$ signal.
We do not see any clear resonant structure in the $\MKeta$ distribution.

The systematic error on the signal yield due to the fitting procedure
is estimated by varying the value of each fixed parameter
by $\pm 1\sigma$ and extracting the new signal yield for each case.
The difference between the background shape for the continuum MC with
and without the LR requirement is taken as an additional error to
the continuum background shape.
We set the normalization of either the generic $\BB$ or rare $B$
backgrounds to zero and to twice its nominal value
to account for its uncertainty.
The changes of the yields for each procedure
are added in quadrature, and are regarded as the systematic error
on the signal yield.
We also calculate a statistical significance for each case,
and regard the smallest value as the significance including the
systematic error.
The result is listed in Table~\ref{tab:summary1}.

\begin{table}
 \begin{center}
  \caption{\label{tab:summary1}%
  Measured signal yields, efficiencies, branching fractions ($\BF$)
  and significances including systematic error ($\mathcal{S}$)
  for $B \to K\eta\gamma$.
  The first and second errors are statistical
  and systematic, respectively.
  Efficiencies include the sub-decay branching fractions.}
  \begin{tabular}{lccccc}
   \hline\hline
   \multicolumn{1}{c}{Mode} & Yield & Efficiency (\%)
   & $\BF$ ($\times 10^{-6}$) & $\mathcal{S}$ \\ \hline
   $B^+ \to K^+ \eta\gamma$ & $\YieldCharged$
   & $\EFFc$ & $\BFketagammaZc$ & $\SNFsystCharged$ \\
   $B^0 \to K^0 \eta\gamma$ & $\YieldNeutral$
   & $\EFFn$ & $\BFketagammaZn$ & $\SNFsystNeutral$ \\
   $B \to K\eta\gamma$ & $\YieldTotal$
   & $\EFFt$ & $\BFketagammaZt$ & $\SNFsystTotal$ \\ \hline\hline
  \end{tabular}
 \end{center}
\end{table}

The signal reconstruction efficiency is estimated using
the MC simulation and is corrected for discrepancies
between data and MC using control samples.
The signal MC has uniform $K\eta$ invariant mass
and $\coshel$ distributions,
where $\thetahel$ is the decay helicity angle
between the kaon momentum and opposite to $B$ momentum 
in the $K\eta$ rest frame.
We find that the efficiency is almost independent of the $K\eta$
invariant mass and $\coshel$.
Table~\ref{tab:summary1} shows
the signal efficiencies and the branching fractions
for each $B \to K\eta\gamma$ mode.
Here, we assume an equal production rate for
$B^0\overline{B}^0$ and $B^+B^-$.
The error on the branching fraction
includes the following systematic uncertainties:
photon detection ($\SYSTphotonSTR$),
tracking ($1.0\%$ to $1.2\%$ per track),
kaon identification ($\SYSTkaonidSTR$),
pion identification ($\SYSTpionidSTR$ per pion),
$\KS$ detection ($\SYSTksSTR$),
$\pi^0$ detection ($\SYSTpizeroSTR$),
$\eta$ detection in $\eta \to \gamma\gamma$ mode ($\SYSTetaSTR$),
$\pi^0/\eta$ veto and LR ($\SYSTlrsystCstr$ and $\SYSTlrsystNstr$ for
charged and neutral modes, respectively),
possible $K\eta$ mass dependence of the efficiency
($\SYSTketamassCstr$ and $\SYSTketamassNstr$ for
charged and neutral modes, respectively),
possible $\coshel$ dependence of the efficiency
($\SYSTcoshelCstr$ and $\SYSTcoshelNstr$ for
charged and neutral modes, respectively),
uncertainty in the $\eta$ branching fraction
($\SYSTsubbfgSTR$ for $\eta \to \gamma\gamma$
and $\SYSTsubbfpSTR$ for $\eta \to \pi^+\pi^-\pi^0$),
and uncertainty in the number of $\BB$ events ($\SYSTNBBstr$).
The systematic errors from the $\pi^0/\eta$ veto and LR requirement
are estimated using
control samples of $B^+ \to \overline{D}{}^0(\to K^+\pi^-\pi^0)\pi^+$ and
$B^0 \to D^-(\to \KS\pi^-\pi^0)\pi^+$ decays,
treating the primary pion
as a high energy photon.

\begin{figure}
 \begin{center}
  \includegraphics[scale=0.62]{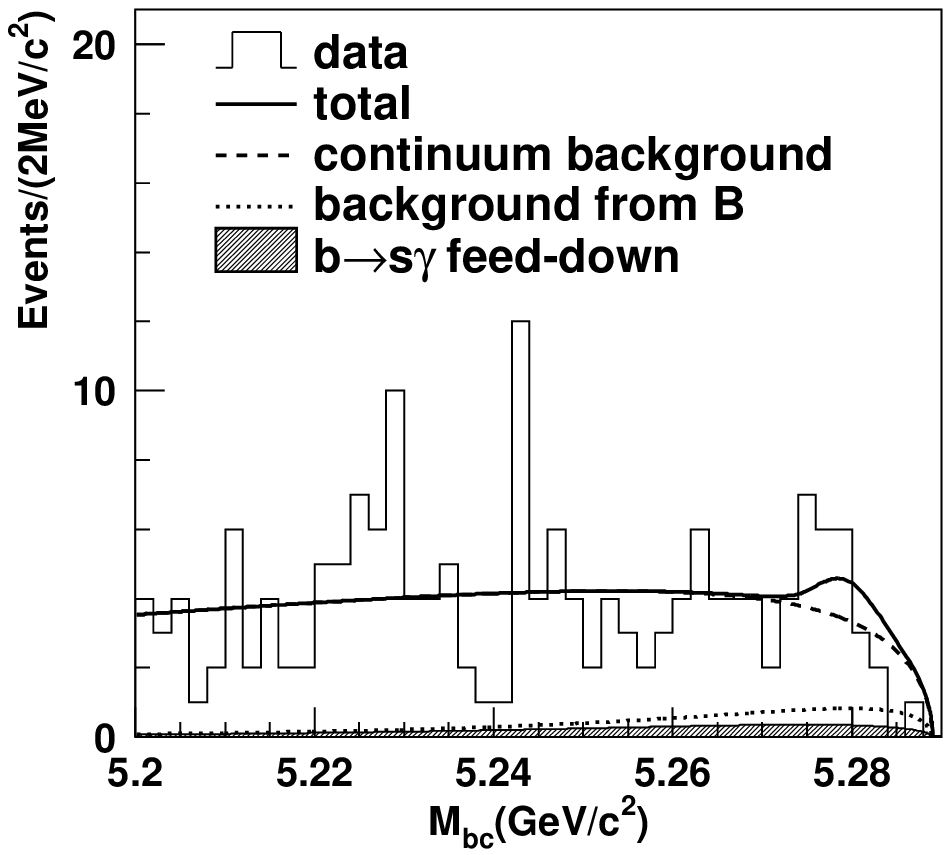}
  \caption{\label{ketagamma_fig4}%
  $\Mbc$ distribution for combined $B \to K\eta\gamma$
  with the $B \to \Kthreest\gamma$ selection.
  Fit results are overlaid.}
 \end{center}
\end{figure}

We search for the decay $B \to \Kthreest\gamma$ by applying
the additional requirements $1.60~\GeV/c^2 < \MKeta < 1.95~\GeV/c^2$
and $|\coshel|<0.2$ or $|\coshel|>0.7$.
The expected $\coshel$ distribution for $B \to \Kthreest\gamma$
is proportional to
$1 - 11 \cos^2\thetahel + 35 \cos^4\thetahel - 25 \cos^6\thetahel$.
The fits to the $\Mbc$ distributions yield
$\YieldTc$, $\YieldTn$ and $\YieldTt$ events
for the charged, neutral and combined modes, respectively.
Here and in the following, we quote statistical and systematic errors
in the first and second position.
The $\Mbc$ distribution and fit result for the combined mode is
shown in Fig.~\ref{ketagamma_fig4}.
We provide only upper limits due to our inability
to distinguish $B \to \Kthreest\gamma$ from non-resonant decays.
The 90\% confidence level upper limit $N$ is calculated from the relation
$\int^{N}_{0} \mathcal{L}(n)dn = 0.9 \int^{\infty}_{0} \mathcal{L}(n)dn$,
where $\mathcal{L}(n)$ is the maximum likelihood
in the $\Mbc$ fit with the signal yield
fixed at $n$. In order to include the systematic
errors from the fitting procedure
in the upper limit for the yield,
the positive systematic error is added to $N$.
The obtained yield upper limits, efficiencies and 
products of branching fractions
$\BF(B \to \Kthreest\gamma) \times \BF(\Kthreest \to K\eta)$
are listed in Table~\ref{tab:summary2}.
Here, the number of $\BB$ events and the reconstruction efficiency are
lowered by $1\sigma$ when we calculate the upper limit
for the branching fractions.
If we assume
$\BF(\Kthreest \to K\eta) = (11 \PM{5}{4})\%$~\cite{Yost:1988ke},
the $90\%$ confidence level limits correspond to
$B \to \Kthreest\gamma$ branching fractions of 
$\BFkthreestgammaULcSBstr$, $\BFkthreestgammaULnSBstr$
and $\BFkthreestgammaULtSBstr$, respectively for charged, neutral
and combined modes, which substantially
improve the limits set by the ARGUS collaboration~\cite{Albrecht:1988ud}.

\begin{table}
 \begin{center}
  \caption{\label{tab:summary2}%
  Measured signal yields, efficiencies and
  products of branching fractions of $B \to \Kthreest\gamma$ 
  and $\Kthreest \to K\eta$ ($\BF \times \BF(K_3^* \to K\eta)$).
  Efficiencies include the sub-decay branching fractions of $\eta$ and
  $K^0$, but not of $\Kthreest$.
  Upper limits are calculated at the $90\%$ confidence level
  and include systematics.}
  \begin{tabular}{lcccc}
   \hline\hline
   \multicolumn{1}{c}{Mode} & Yield & Efficiency (\%)
   & $\BF \times \BF(K_3^* \to K\eta)$ ($\times 10^{-6}$)
   \\ \hline
   $B^+ \to \Kthreest^+\gamma$ & $< \YieldULTc$
   & $\EFFTc$ & $< \BFkthreestgammaULc$ \\
   $B^0 \to \Kthreest^0\gamma$ & $< \YieldULTn$
   & $\EFFTn$ & $< \BFkthreestgammaULn$ \\
   $B \to \Kthreest\gamma$ & $< \YieldULTt$
   & $\EFFTt$ & $< \BFkthreestgammaULt$ \\ \hline\hline
  \end{tabular}
 \end{center}
\end{table}

\begin{figure}
 \begin{center}
  \includegraphics[scale=0.62]{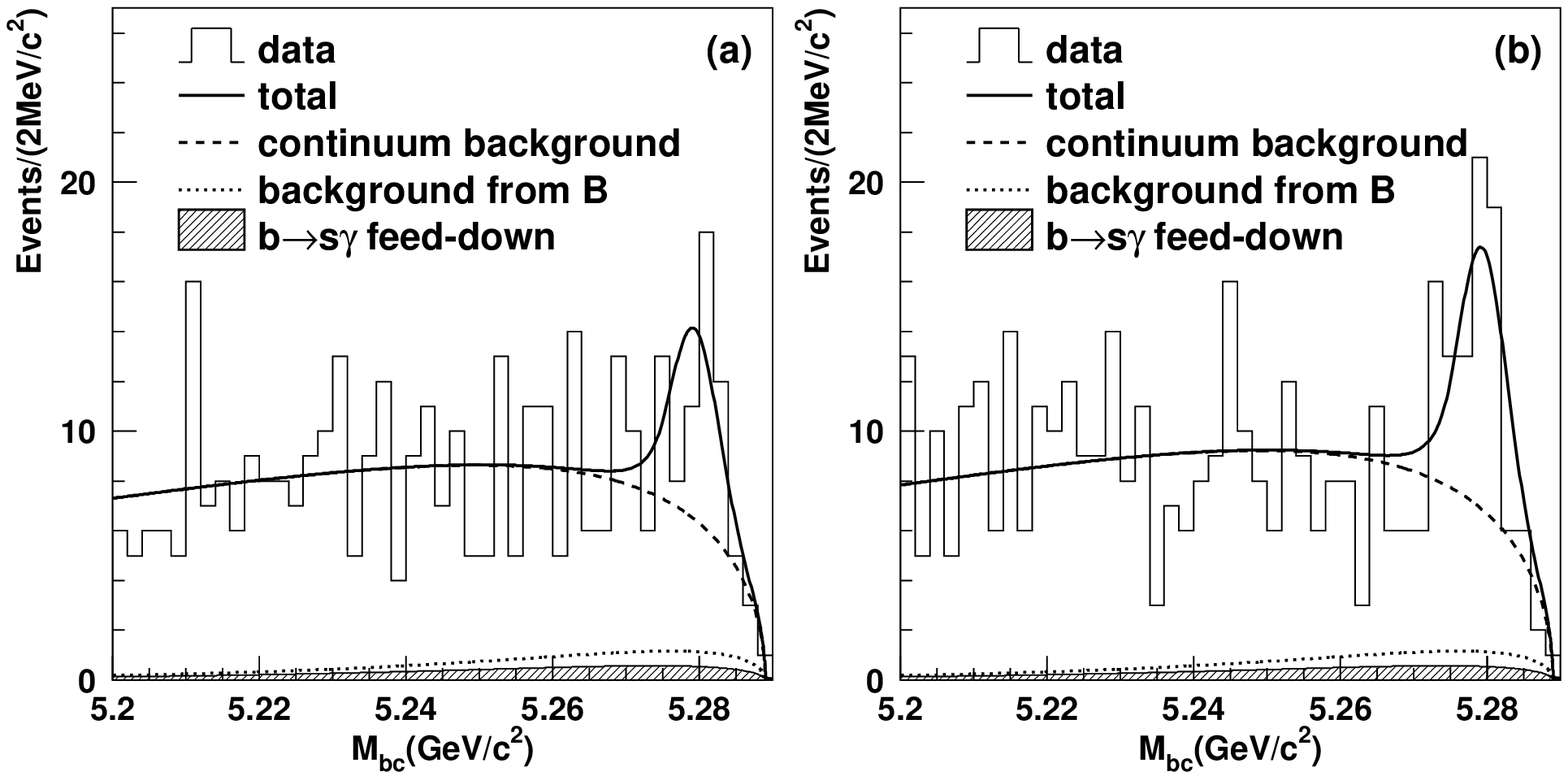}
  \caption{\label{ketagamma_fig5}%
  $\Mbc$ distributions for
  (a) negative charged $B^- \to K^-\eta\gamma$,
  (b) positive charged $B^+ \to K^+\eta\gamma$.
  Fit results are overlaid.}
 \end{center}
\end{figure}

Some extensions of the SM predict a large $\mathit{CP}$ asymmetry
in the $\bsgamma$ process~\cite{ACP-predictions}.
We measure the partial rate asymmetry
$\ACP = ( 1 / ( 1 - 2w ) ) ( N_- - N_+ ) / ( N_- + N_+ )$
for $B^+ \to K^+\eta\gamma$, where
$N_{\mp}$ is the signal yield for $B^{\mp} \to K^{\mp}\eta\gamma$ and
$w$ is the probability that a signal event is reconstructed with the 
wrong kaon (and hence $B$) charge. This probability is found to be less
than $1\%$ in our signal MC sample,
and hence we ignore its negligible effect on $\ACP$.
$N_{\mp}$ is obtained by fitting separately the $\Mbc$ distributions
for the negatively and positively charged modes shown
in Fig.~\ref{ketagamma_fig5}.
We find $N_- = \YieldZcMinus$ and $N_+ = \YieldZcPlus$.
The systematic error on $\ACP$ consists of the following contributions.
The error from the fitting procedure
is estimated to be $\AcpSystFit$ by varying each fixed parameter one by one,
and extracting $\ACP$ for each procedure, in the same way as before.
Here, we assume no asymmetry for the generic $\BB$ background,
but allow $100\%$ asymmetry for the rare $B$
and $6\%$ asymmetry for $\bsgamma$~\cite{Nishida:2003yw}.
The error from the overall detector bias is studied with the
$B^0 \to D^-(K^-\pi^+\pi^0)\pi^+$ control sample
and is found to be $\AcpSystDet$.
By adding these errors and the possible asymmetry
in kaon identification ($\AcpSystPID$) in quadrature,
we obtain $\ACP = \AcpZc$.

In conclusion, we observe the decay mode $B^+ \to K^+\eta\gamma$
and find the first evidence of $B^0 \to K^0\eta\gamma$.
The branching fraction and partial rate asymmetry
of $B^+ \to K^+\eta\gamma$ are measured to be
$(\BFketagammaZc) \times 10^{-6}$ and $\AcpZc$
for $\MKeta < 2.4~\GeV/c^2$.
The branching fraction of $B^0 \to K^0\eta\gamma$ is 
measured to be $(\BFketagammaZn) \times 10^{-6}$.
We also search for $B \to \Kthreest\gamma$, but find no evidence.
Although the signal yield for $B^0 \to \KS\eta\gamma$ is small,
this mode can be used in the near future to study
time-dependent $\CP$ asymmetries in radiative $B$ decays
and to search for new physics.


We thank the KEKB group for the excellent operation of the
accelerator, the KEK cryogenics group for the efficient
operation of the solenoid, and the KEK computer group and
the National Institute of Informatics for valuable computing
and Super-SINET network support. We acknowledge support from
the Ministry of Education, Culture, Sports, Science, and
Technology of Japan and the Japan Society for the Promotion
of Science (JSPS); the Australian Research Council and the
Australian Department of Education, Science and Training;
the National Science Foundation of China under contract
No.~10175071; the Department of Science and Technology of
India; the BK21 program of the Ministry of Education of
Korea and the CHEP SRC program of the Korea Science and
Engineering Foundation; the Polish State Committee for
Scientific Research under contract No.~2P03B 01324; the
Ministry of Science and Technology of the Russian
Federation; the Ministry of Education, Science and Sport of
the Republic of Slovenia;  the Swiss National Science Foundation;
the National Science Council and
the Ministry of Education of Taiwan; and the U.S.\
Department of Energy.
S.N. is supported by KAKENHI(16740153) by JSPS.


\begin{thebibliography}{99}
 \bibitem{Eidelman:2004wy}
	 S. Eidelman \textit{et al.} (Particle Data Group),
	 Phys. Lett. B 592 (2004) 1.
 \bibitem{Coan:1999kh}
         T. E. Coan \textit{et al.} (CLEO Collaboration),
         Phys. Rev. Lett. 84 (2000) 5283.
 \bibitem{Aubert:2001me-Nakao:2004th}
	 B.~Aubert \textit{et al.} (BABAR Collaboration),
	 Phys. Rev. Lett. 88 (2002) 101805;
	 M.~Nakao \textit{et al.} (Belle Collaboration),
	 Phys. Rev. D 69 (2004) 112001.
 \bibitem{Nishida:2002me}
	 S. Nishida \textit{et al.} (Belle Collaboration),
	 Phys. Rev. Lett. 89 (2002) 231801.
 \bibitem{Drutskoy:2003xh}
	 A. Drutskoy \textit{et al.} (Belle Collaboration),
	 Phys. Rev. Lett. 92 (2003) 051801.
 \bibitem{Kaneko:2002mr}
	 J. Kaneko \textit{et al.} (Belle Collaboration),
	 Phys. Rev. Lett. 90 (2003) 021801.
 \bibitem{TCPV-bsgamma}
	 D.~Atwood, M.~Gronau and A.~Soni,
	 Phys. Rev. Lett. 79 (1997) 185;
	 D.~Atwood \textit{et al.}, hep-ph/0410036.
 \bibitem{Aston:1987ey}
	 D. Aston \textit{et al.} (LASS Collaboration),
	 Phys. Lett. B 201 (1988) 169.
 \bibitem{Mori:2000cg}
	 A. Abashian \textit{et al.} (Belle Collaboration),
	 Nucl. Instrum. Meth. A 479 (2002) 117.
 \bibitem{KEKB:NIM}
	 S.~Kurokawa and E.~Kikutani,
	 Nucl. Instrum. Meth., A 499 (2003) 1,
	 and other papers included in this volume.
 \bibitem{Ushiroda} Y. Ushiroda (Belle SVD2 Group),
	 Nucl. Instr. and Meth. A 511 (2003) 6.
 \bibitem{Fisher:1936et}
	 R. A. Fisher, Annals Eugen. 7 (1936) 179.
 \bibitem{Fox:1978vu}
	 G. C. Fox and S. Wolfram, Phys. Rev. Lett.
	 41 (1978) 1581.
 \bibitem{Crystal-Ball}
	 T. Skwarnicki, Ph.D. Thesis, Institute for Nuclear Physics,
	 Krakow 1986; DESY Internal Report, DESY F31-86-02 (1986).
 \bibitem{Albrecht:1990am}
 	 H. Albrecht \textit{et al.} (ARGUS Collaboration),
 	 Phys. Lett. B 241 (1990) 278.
 \bibitem{Sjostrand:1994yb}
	 T. Sj{\"o}strand,
	 Comput. Phys. Commun. 82 (1994) 74.
 \bibitem{Kagan:1998ym}
	 A.L. Kagan and M. Neubert,
	 Eur. Phys. J. C 7 (1999) 5.
 \bibitem{Yost:1988ke}
	 G. P. Yost \textit{et al.} (Particle Data Group),
	 Phys. Lett. B 204 (1988) 1;
	 the value of $\BF(\Kthreest \to K\eta)$ quoted in
	 Ref.~\cite{Eidelman:2004wy} seems to be invalid
	 since it does not take into account a study of the
	 $\Kthreest \to K \eta$ decay mode in Ref.~\cite{Aston:1987ey}.
 \bibitem{Albrecht:1988ud}
	 H. Albrecht \textit{et al.} (ARGUS Collaboration),
	 Phys. Lett. B 210 (1988) 258.
 \bibitem{ACP-predictions}
         For example,
         K. Kiers, A. Soni and G.-H. Wu, Phys. Rev. D 62 (2000) 116004;
         A. L. Kagan and M. Neubert, Phys. Rev. D 58 (1998) 094012;
         S. Baek and P. Ko, Phys. Rev. Lett. 83 (1998) 488.
 \bibitem{Nishida:2003yw}
	 S. Nishida \textit{et al.} (Belle Collaboration),
	 Phys. Rev. Lett. 93 (2004) 031803.
\end{thebibliography}
\end{document}